\documentstyle[preprint,revtex]{aps}
\begin{document}
\begin{title}
Instantons And Baryon Mass Splittings in the MIT Bag Model
\end{title}
\author{D. Klabu\v car}
\begin{instit}
Physics Department of Zagreb University, P.O.B. 162, 41001 Zagreb,
Croatia
\end{instit}
\begin{abstract}
The contribution of instanton-induced effective inter-quark interactions
to the baryon mass splittings was considered in the bag model. It is
found that results are different from those obtained in the constituent
quark model where the instanton effects are like those from one-gluon
exchange. This is because in the context of the bag model calculation
the one-body instanton-induced interaction has to be included.
\vskip 1.2cm
12.38.Lg; 12.40.Aa; 14.20.-c
\end{abstract}
\end{document}

\def\huh{\hbox{\vrule width 2pt height 8pt depth 2pt}}
\def\Bbb#1{{\bf #1}}
\def\eqnum#1{\eqno (#1)}
\def\fnote#1{\footnote}
\def\blacksquare{\hbox{\vrule width 4pt height 4pt depth 0pt}}
\def\square{\hbox{\vrule\vbox{\hrule\phantom{o}\hrule}\vrule}}
\def\cwdash{\relbar\joinrel}
\def\cwleft{0}
\def\cwright{1}
\def\cwcenter{2}
\def\cwtable{3}
\def\cwfull{4}
\def\cwoneandhalf{0}
\def\cwdouble{1}
\def\cwtrible{2}
\def\cwsingle{3}
\def\cwleftpar#1#2{\leftskip #1 \rightskip #2 plus 1fill}
\def\cwrightpar#1#2{\leftskip #1 plus 1fill \rightskip #2}
\def\cwcenterpar#1#2{\leftskip #1 plus 1fill \rightskip #2 plus 1fill}
\def\cwfullpar#1#2{\leftskip#1\rightskip#2}
\def\cwindent#1{\noindent \hangindent #1\hangafter -1}
\def\cwoutdent#1#2{\llap{\hbox to #1{#2 \hss}}\ignorespaces}
\def\cwparbegin#1#2#3#4#5{
	\ifcase #1 \cwleftpar{#2}{#3}
	\or \cwrightpar{#2}{#3}
	\or \cwcenterpar{#2}{#3}
	\else \cwfullpar{#2}{#3}\fi
	\ifcase #4 \baselineskip = 1.5\baselineskip
	\or \baselineskip = 2\baselineskip
	\or \baselineskip = 3\baselineskip
	\else \baselineskip = 1\baselineskip\fi
	\ifdim #5 > 0in \else \noindent \fi
	\noindent\ignorespaces}
\def\tj0{ {{\widetilde  j}_0} }
\def\tj1{ {{\widetilde  j}_1} }
\documentstyle[12pt]{article}
\begin{document}
\leftline{{\bf   I.  Introduction and motivation} }
\medskip
{\cwparbegin{\cwfull}{0.00in}{0.00in}{\cwdouble}{0.38in}
\cwindent{0.38in} QCD instantons are supposed to have interesting consequences
for the structure of hadrons.  Nevertheless, their role is difficult to
disentangle from
the other effects and the ways of embedding the instanton-related effects in
the
description of the hadronic structure can still be somewhat ambiguous.
\par}
{\cwparbegin{\cwfull}{0.00in}{0.00in}{\cwdouble}{0.38in}
\cwindent{0.38in}As an example, let us consider the role of instantons in the
baryon mass
splittings. Shuryak and Rosner [1] studied them in the constituent quark model
and concluded that an effective instanton two-body interaction provides as
satisfactory a description of mass splittings in the octet and decuplet baryons
as the more conventional picture based on hyperfine interaction due to
one-gluon
exchange. So, in the nonrelativistic constituent quark model, instantons and
one-gluon exchange are doing essentially the same thing as far as the mass
spectrum is concerned. While this makes it problematic to disentangle which
part
in the mass shifts comes from one-gluon exchange and which from instanton
effects, it also opens some attractive possibilities of refining the models of
hadronic structure.
\par}
{\cwparbegin{\cwfull}{0.00in}{0.00in}{\cwdouble}{0.38in}
\cwindent{0.38in}To illustrate the above statement let us consider the MIT
quark bag [2].
This relativistic model is in some sense complementary to the constituent
quark
model and it is therefore important to see if the effects of instantons in
these two
models are mutually consistent. Moreover, if instanton effects in the bag model
turn out to be similar to the effects of one-gluon exchange, a possibility
opens
for a much needed refinement of the bag model, i.e. the reduction of the strong
coupling constant to a value which would be truly consistent with perturbation
theory. Namely, the bag model is a marriage of two opposite regimes: it
captures
the long-distance, confining effects of QCD by postulating a confining
boundary. On the other hand, inside this boundary quarks are supposed to
interact by perturbative QCD (supposedly saturated by  one-gluon exchange)
which
describes the physics of small inter-quark separations.
Thus,  it is in fact not surprising that  MIT bag model fits
always require too large  strong coupling constant $\alpha_c$,
since this "perturbative" one-gluon exchange is forced to
account for all non-confining quark interactions inside the cavity (the
confinement part is summarized by impregnable boundary) where inter-quark
separations can be as high as $2R_{\hbox{bag}}\simeq  2$ fm!
(The situation is somewhat different in some more elaborate bag models, like in
some
variants of  the chiral bag model [3]. E.g. in the chiral little bag model
[4], the
inner quark core is squeezed to $R_{bag}\simeq 0.5 $ fm by a meson soliton
outside and such a
configuration requires smaller $\alpha _{c}$ inside the bag. However, in this
model the
meson soliton pretty much sums up the long- and intermediate-range
nonperturbative
gluon effects, and in this work we want to see if - and to what extent -
these effects can be described by instantons. That is why we stick with the
simplest, MIT bag model [2] and do not consider bag models refined by, e.g.,
meson solitons.)
\par}
{\cwparbegin{\cwfull}{0.00in}{0.00in}{\cwdouble}{0.38in}
\cwindent{0.38in}Now, although the effective instanton-induced quark
interactions are
usually (at least in model calculations) considered in the local approximation,
they stem from instantons which are nonperturbative structures of an average
spatial scale of about $1/600$ MeV $\approx  1/3 $ fm. They are therefore of
just the right
scale to help capture the intermediate-range QCD effects. (It is by now quite
certain that they are not responsible for confinement [5], as thought
previously.)
If they contribute to the mass shifts in the same direction as  one-gluon
exchange,
the latter is freed from a part of its task in producing the mass shifts and
can
be reduced in magnitude by reducing the value of the strong coupling constant.
Such smaller coupling constant would be more appropriate for the short-distance
physics and would hopefully be truly perturbative.
\par}
{\cwparbegin{\cwfull}{0.00in}{0.00in}{\cwdouble}{0.38in}
\cwindent{0.38in}Having so listed the motivation for checking what happens if
one tries to
introduce instantons in the MIT bag model, let us now explain how we go about
doing this.
\par}
\medskip
\medskip
{\cwparbegin{\cwfull}{0.00in}{0.00in}{\cwdouble}{0.00in}
{\bf    II.  Incorporation of the instanton-induced interaction}
\par}
{\cwparbegin{\cwfull}{0.00in}{0.00in}{\cwoneandhalf}{0.00in}
{\bf in the MIT bag model}
\par}
{\cwparbegin{\cwfull}{0.00in}{0.00in}{\cwdouble}{0.38in}
\cwindent{0.38in}We shall consider the vacuum averaged version of the
instanton-induced
effective interaction between quarks derived for the case of instanton liquid
by
Nowak et al.  [6] , but transformed to $x$-space. It is essentially the same as
Shifman-Veinstein-Zakharov (SVZ) interaction of Ref. [7]. It is convenient to
separate it in "zero-body", one-body, two-body and three-body pieces,
\par}
\medskip
{\cwparbegin{\cwtable}{0.00in}{0.00in}{\cwoneandhalf}{0.38in}
\cwindent{0.38in}\halign{\hfil#\hfil\hskip 3em &\hfil#\hskip 3em \cr
\hskip 3.2cm ${\cal L}_{I}= {\cal L}^{I}_{0}+ {\cal L}^{I}_{1}
+ {\cal L}^{I}_{2}+ {\cal L}^{I}_{3} .\qquad\qquad\qquad $ &(1)\cr
}\par}
\medskip
{\cwparbegin{\cwfull}{0.00in}{0.00in}{\cwdouble}{0.00in}
What we  termed "zero-body" part ${\cal L}^{I}_{0}$  is simply
a constant, a C-number combination of current
quark masses $m_{u}, m_{d}, m_{s}$,  instanton density $n$ and
instanton average size $\rho $, {\it i.e.}
it does not contain any quark field operators.
It corresponds to a constant energy density which
is the same for any state (just like the bag constant $B$) so that its
matrix element for a given hadron $H$ is  simply a part of the volume energy,
i.e. it  is of the same form as, and should contribute to,
the usual volume energy
$B (4\pi/3) R_H^3$. That is,
the ${\cal L}^I_0$ contribution amounts
to the renormalization of $B$. However, the bag constant is in practice
not obtained by calculating various contributions to $B$ and then
summing them all up, but it is determined as a whole
by a phenomenological fit to the hadron spectrum. Thus $B$ contains all
contributions, including the one due to ${\cal L}^I_0$, so that the explicit
inclusion of ${\cal L}^I_0$ in the instanton-induced mass shifts would
be double-counting. Of course, the fitted value of $B$
will change after the relevant parts of ${\cal L}_I$
are included in the calculation of hadron masses,
namely these parts containing the quark field operators:
\par}
\medskip
\medskip
\medskip
{\cwparbegin{\cwtable}{0.00in}{0.00in}{\cwoneandhalf}{0.00in}
\halign{\hfil#\hfil\hskip 3em &\hfil#\hskip 3em \cr
${\cal L}^{I}_{1}= -n\left(\begin{array}{c}
{4\pi { } ^{2}\over 3}{\rho ^{3}}\end{array}\right)
\lbrace{\cal F}_{u}\bar{u}_{R}u_{L}
+ (u\leftarrow\joinrel \rightarrow d)
+ (u\leftarrow\joinrel \rightarrow s) \rbrace
+ (R \leftarrow\joinrel \rightarrow L)$,&(2)\cr
}\par}
\medskip
\medskip
\medskip
{\cwparbegin{\cwtable}{0.00in}{0.00in}{\cwsingle}{0.00in}
\halign{\hfil#\hfil\hskip 3em &\hfil#\hskip 3em \cr
${\cal L}^{I}_{2}= -n\left(\begin{array}{c}
{4\pi { } ^{2}\over 3}{\rho ^{3}}\end{array}\right) ^{2}
\lbrace {\cal F}_{u}{\cal F}_{d}
[(\bar{u}_{R}u_{L})(\bar{d}_{R}d_{L})+ {3\over 32}
(\bar{u}_{R}\lambda ^{a}u_{L}\bar{d}_{R}\lambda ^{a}d_{L}$& \qquad (3)\cr
}\par}
\medskip
\medskip
\medskip
{\cwparbegin{\cwfull}{0.92in}{0.00in}{\cwsingle}{0.00in}
$\!\!\!\!\!\!\!\!\!\!\! \!\!\!\!\!\!\!
- {3\over 4} \bar{u}_{R}\sigma _{\mu \nu }\lambda ^{a}u_{L}
\bar{d}_{L}\sigma ^{\mu \nu }\lambda ^{a}d_{L})]
+ (u\leftarrow\joinrel \rightarrow s)
+ (d\leftarrow\joinrel \rightarrow s) \rbrace
+ (R\leftarrow\joinrel \rightarrow L)$,
\par}
\medskip
\medskip
\medskip
{\cwparbegin{\cwtable}{0.00in}{0.00in}{\cwsingle}{0.00in}
\halign{#\hfil\hskip 3em \cr
\cr
}\par}
\medskip
{\cwparbegin{\cwtable}{0.00in}{0.00in}{\cwsingle}{0.00in}
\halign{\hfil#\hfil\hskip 3em &\hfil#\hskip 3em \cr
${\cal L}^{I}_{3}= -n\left(\begin{array}{c}
{4\pi { } ^{2}\over 3}{\rho ^{3}}\end{array}\right) ^{3}
\, {\cal F}_{u}\, {\cal F}_{d}\, {\cal F}_{s}\, {1\over 3!}\, {1\over
{N_{c}(N^{2}_{c} - 1)}} \,
\epsilon_{f_{1} f_{2} f_3} \, \epsilon _{g_{1} g_2 g_3}\lbrace\begin{array}{l}
\left[\begin{array}{c}
1 - {3\over 2(N_{c}+2)}\end{array}\right] \end{array} $ &\cr
}\par}
\medskip
\medskip
\medskip
{\cwparbegin{\cwfull}{0.54in}{0.00in}{\cwsingle}{-0.54in}
\cwoutdent{0.54in}{}
$\!\!\times (\bar{q}^{f_{1}}_{R}q^{g_{1}}_{L})
(\bar{q}^{f_{2}}_{R}q^{g_{2}}_{L})(\bar{q}^{f_{3}}_{R}q^{g_{3}}_{L}) +
{8\over 3(N_{c}+3)} (\bar{q}^{f_{1}}_{R}q^{g_{1}}_{L})
(\bar{q}^{f_{2}}_{R}\sigma_{\mu \nu }q^{g_{2}}_{L})
(\bar{q}^{f_{3}}_{R}\sigma^{\mu \nu }q^{g_{3}}_{L}) \rbrace$    \qquad (4)
\par}
\medskip
\medskip
\medskip
{\cwparbegin{\cwfull}{0.54in}{0.00in}{\cwsingle}{-0.54in}
\cwoutdent{0.54in}{} $
+ (R\leftarrow\joinrel \rightarrow L).$
\par}
\medskip
{\cwparbegin{\cwfull}{0.00in}{0.00in}{\cwdouble}{0.00in}
The left (and right) projected components are standard, {\it e.g.}
\par}
\medskip
{\cwparbegin{\cwtable}{0.00in}{0.00in}{\cwdouble}{0.00in}
\halign{\hfil#\hfil\hskip 3em &\hfil#\hskip 3em \cr
\hskip 3.2cm $u_{L\atop R} =
\gamma_{\pm}u \equiv  {1\over 2} (1\pm \gamma_{5}) u$,&(5)\cr
}\par}
\medskip
{\cwparbegin{\cwfull}{0.00in}{0.00in}{\cwdouble}{0.00in}
etc. ${\cal F}_{f}$'s are the characteristic factors (corresponding to inverse
effective
masses) composed of current quark masses $m_{f}$  $(f = u,d,s)$, average
instanton
size $\rho $ and quark condensate $<0\mid \bar{q} q\mid 0> =
(- 240$ MeV$)^3$, {\it e.g.}
\par}
\medskip
{\cwparbegin{\cwtable}{0.00in}{0.00in}{\cwdouble}{0.00in}
\halign{\hfil#\hfil\hskip 3em &\hfil#\hskip 3em \cr
\hskip  3.2cm ${\cal F}_{u}\equiv \left[\begin{array}{c}
m_{u}\rho  - {2\pi { } ^{2}\over 3} \rho ^{3}
<0\mid \bar{q} q\mid 0>\end{array}\right]^{-1} $ &(6)\cr
}\par}
\medskip
{\cwparbegin{\cwfull}{0.00in}{0.00in}{\cwdouble}{0.00in}
and analogously for $d$ and $s$ flavors. In the three-body interaction ${\cal
L}^{I}_{3}$, the
indices $f_{i}, g_{i}$  $(i = 1,2,3)$ run over flavors $u, d$ and $s$. (E.g.
$g_{2}=u$ means
$q^{g_{2}}_{L} \equiv  u_{L})$. Summation over repeated indices is understood,
so that the first term of  ${{\cal L}_3^I}$ (which leads in ${1/N_{c}} $  and
does
not contain $\sigma _{\mu \nu })$ is simply the quark determinant. The
three-body
interaction indeed looks surprisingly simpler than one would expect  from,
e.g., SVZ
version [7]. This remarkable simplification has been detailed by Nowak   [8]
who
Fierzed away otherwise very complex color structures in the three-body piece
SVZ
interaction [7] and lumped it in simple prefactors containing $N_{c}$.
\par}
{\cwparbegin{\cwfull}{0.00in}{0.00in}{\cwdouble}{0.38in}
\cwindent{0.38in}However, as each piece of ${\cal L}^{I}_{3}$ always contains
all
three flavors, it would in our calculation contribute only for $\Lambda$
baryon,
which we therefore skip and thus avoid the need to calculate the ${\cal
L}^{I}_{3}$
contribution to baryon mass shifts.
($\Sigma^0$ also contains all three flavors and may have a nonvanishing
${\cal L}^I_3$-contribution. The latter, however, must vanish in the
isospin limit to yield the mass shift equal to the ones of $\Sigma^\pm$,
the isospin partners of $\Sigma^0$.)
\par}
{\cwparbegin{\cwfull}{0.00in}{0.00in}{\cwdouble}{0.38in}
\cwindent{0.38in}The total instanton-induced mass shift of any baryon $\mid B>
($except $\Lambda$  where also ${\cal L}^{I}_{3}$ can contribute) is thus just
\par}
\medskip
{\cwparbegin{\cwtable}{0.00in}{0.00in}{\cwdouble}{0.38in}
\cwindent{0.38in}\halign{\hfil#\hfil\hskip 3em &\hfil#\hskip 3em \cr
\hskip 2.5cm $E^{B}_{I}= \Delta M^{(1)}_{B}+ \Delta M^{(2)}_{B}=
<B\mid : -{\cal L}^{I}_{1}- {\cal L}^{I}_{2}: \mid B>$&(7)\cr
}\par}
\medskip
{\cwparbegin{\cwfull}{0.00in}{0.00in}{\cwdouble}{0.00in}
It is appropriate to give $\Delta M^{I}_{B}$ as the normal ordered interaction
sandwiched
between ordinary bag states $\mid B>$ composed of valence quarks only, so that
there
is no vacuum contributions, because we use the vacuum averaged
instanton-induced
interaction which already includes all relevant vacuum contributions.
\par}
{\cwparbegin{\cwfull}{0.00in}{0.00in}{\cwdouble}{0.38in}
\cwindent{0.38in}The instanton-induced mass shifts have been already studied by
Kochelev [9] for nonstrange quark bags, but there are crucial differences
between his
calculation and ours. First, Kochelev did not include the one-body term but
only
the two-body term ${\cal L}^{I}_{2}$. Dropping of the one-body term would be
justified in the
constituent quark model, where one uses quark masses already "dressed" by QCD,
so that the self-mass part of the instanton effects is already included in the
constituent quark mass parameters. However, we use the current quark masses, as
appropriate in the bag model, so that the effect of ${\cal L}^{I}_{1}$ should
be included. We
find this contribution absolutely crucial, being not only larger than ${\cal
L}^{I}_{2}$
contribution, but also of the opposite sign.
\par}
{\cwparbegin{\cwfull}{0.00in}{0.00in}{\cwdouble}{0.38in}
\cwindent{0.38in}The second difference is that we employ the MIT bag model,
whereas Kochelev
used his own variant of the bag, called the chiral bag model and developed in
Ref. [10]  so that it be in agreement with the sum rules. (However, it differs
substantially from the more standard versions of the chiral bag, e.g. [3], [4]
and [11].)
\par}
{\cwparbegin{\cwfull}{0.00in}{0.00in}{\cwdouble}{0.38in}
\cwindent{0.38in}Moreover, Kochelev assumed the instanton density $n$ to be
equal to the
density in the nonperturbative QCD vacuum between the bag radius $R$ and some
$R_{ch} \simeq  {2\over 3} R$ and then falling to zero. (I.e., in this respect
it is like some "bag
with skin".)
\par}
{\cwparbegin{\cwfull}{0.00in}{0.00in}{\cwdouble}{0.38in}
\cwindent{0.38in}On the other hand, in this work we want to stick firmly with
the original
MIT bag model. The only modification is the inclusion of the instanton-induced
interaction. It is hoped it would describe
intermediate-range ($\approx  {1\over 3}$ fm) physics
anywhere quarks can go within the bag, just as one-gluon exchange should take
care of the short-distance quark-quark interactions anywhere inside the
impenetrable cavity. It is thus appropriate to assume a constant (though as yet
undetermined) instanton density $n = n_{0}$ throughout the bag,
just as ${\alpha}_{c}$ has the
same value everywhere. Clearly, $n$ inside should be significantly smaller
than instanton density  $n_c$
in the true, nonperturbative QCD vacuum  -  otherwise, the vacuum inside the
bag would depart very much from the trivial perturbative vacuum and would start
looking more and more like the nonperturbative one. Already in Ref. [12],
Shuryak
argued on general grounds that instanton density should be substantially
depleted inside  a quark bag.  In the present framework, this value of $n$
appropriate for
the interior of the MIT bag, should come out as a result of our model
calculation. Thus, our
$n$ also looks like a step function, but it falls from the true vacuum value
$( n = n_c\approx   8  \cdot 10^{-4} $  GeV$^4$ [12] ) to a smaller
(but in principle nonzero) value to be determined below.
\par}
\medskip
{\cwparbegin{\cwfull}{0.00in}{0.00in}{\cwdouble}{0.00in}
{\bf  { III.    Instantons and the $p-n$ mass difference }}
\par}
{\cwparbegin{\cwfull}{0.00in}{0.00in}{\cwdouble}{0.38in}
\cwindent{0.38in}If we go beyond isospin symmetry and take $u$ and $d$ quark
masses to be
different, $m_{u}\neq m_{d}$, Eq. (7) yields a nonvanishing instanton
contribution to the
proton-neutron $(p-n)$ mass difference,
\par}
{\cwparbegin{\cwtable}{0.00in}{0.00in}{\cwdouble}{0.38in}
\cwindent{0.38in}\halign{\hfil#\hfil\hskip 3em &\hfil#\hskip 3em \cr
\hskip 3.2cm $\Delta E^{pn}_{I}=$ $\widetilde n$ ${4\pi { } ^{2}\over 3} \rho
^{-1}
[{\cal F}_{u}I_{u}- {\cal F}_{d}I_{d}] $,&(8)\cr
}\par}
\medskip
\medskip
{\cwparbegin{\cwtable}{0.00in}{0.00in}{\cwdouble}{0.00in}
\halign{\hfil#\hfil\hskip 3em &\hfil#\hskip 3em \cr
\hskip 3.2cm $I_{f}= \int \bar{\Psi }_{f}\Psi _{f}d^{3}r,  \qquad   f =
u,d,s$,&(9)\cr
}\par}
{\cwparbegin{\cwfull}{0.00in}{0.00in}{\cwdouble}{0.00in}
where $\Psi _{f}$ is the usual ground state wavefunction (given e.g. in Ref.
[13]) of a
bagged quark of mass $m_{f}$. $\widetilde n$ is the "dimensionless instanton
density"
obtained by expressing $n$ in units of the inverse average instanton size $\rho
$,
$ n \equiv  \widetilde n\rho ^{-4}$.  We shall
consistently use the commonly accepted value $\rho  = 1/600$ MeV $\simeq  1/3 $
fm  [14].
\par}
{\cwparbegin{\cwfull}{0.00in}{0.00in}{\cwdouble}{0.38in}
\cwindent{0.38in}The proton-neutron mass difference in the MIT bag model
was studied by Chodos and Thorn [15], then Desphande et al.  [16]
and later by,  e.g., Bickerstaff and Thomas  [17].  In order to make
a consistent comparison with and usage of their results  [17], we give $\Delta
E^{pn}_{I}$ for
the quark masses they used, $m_{u}= 7.86$ MeV and $m_{d}= 12.14$ MeV:
\par}
\medskip
{\cwparbegin{\cwtable}{0.00in}{0.00in}{\cwdouble}{0.38in}
\cwindent{0.38in}\halign{\hfil#\hfil\hskip 3em &\hfil#\hskip 3em \cr
\hskip 3.2cm $\Delta E^{pn}_{I}= {\widetilde n}\cdot 83.137$ MeV. &(10)\cr
}\par}
{\cwparbegin{\cwfull}{0.00in}{0.00in}{\cwdouble}{0.00in}
(The value that $\widetilde n$ may take will be determined below.)
\par}
{\cwparbegin{\cwfull}{0.00in}{0.00in}{\cwdouble}{0.38in}
\cwindent{0.38in}It turns out, however (see, e.g., Ref.  [18]) that the
contributions to $p-n$
difference from the quark kinetic energy $E_{K}$, color-magnetic energy due to
one-gluon exchange $E_{c}$, and the electromagnetic energy $E_{EM}$, are
essentially
sufficient to fit the whole $p-n$ mass difference.  {\it I.e.}, denoting these
contributions by $\Delta E_{X}, (X=K,c,EM,I)$,
\par}
\medskip
{\cwparbegin{\cwtable}{0.00in}{0.00in}{\cwdouble}{0.38in}
\cwindent{0.38in}\halign{\hfil#\hfil\hskip 3em &\hfil#\hskip 3em \cr
\hskip 2.8cm $\Delta E_{pn}= \Delta E^{pn}_{K}+ \Delta E^{pn}_{c}
+ \Delta E^{pn}_{EM}  \qquad\qquad\qquad $  &(11)\cr
}\par}
\medskip
{\cwparbegin{\cwtable}{0.00in}{0.00in}{\cwdouble}{0.38in}
\cwindent{0.38in}\halign{\hfil#\hfil\hskip 3em &\hfil#\hskip 3em \cr
\hskip 2.8cm \hskip 1.2cm $= (-2.09+0.30+0.50)$ MeV $= -1.29$ MeV,\cr
}\par}
{\cwparbegin{\cwfull}{0.00in}{0.00in}{\cwdouble}{0.38in}
and this is equal to the experimental value of $\Delta E_{pn}$.
How can then the instanton contribution $\Delta E^{pn}_{I}$ fit in?  One
possibility is that
phenomenology tells us via $\Delta E_{pn}$ that the instanton density  inside
the bag must
be very small indeed,  $\widetilde n \le  10^{-3}$. The other, more attractive
possibility is that
instanton effects may allow the reduction of the model parameter playing the
role of the strong coupling constant $\alpha _{c}$. Since $\Delta E^{pn}_{c}$
(as well as $E_{c}$) is proportional to $\alpha _{c}$ and $\Delta E^{pn}_{c}$
has the same sign as the instanton contribution $\Delta E^{pn}_{I}$, it is
tempting
to assume that $\Delta E^{pn}_{c}$ can in fact be reduced by reducing $\alpha
_{c}$
to a value acceptable for perturbation theory, while the decrease in $\Delta
E^{pn}_{c}$ would
be compensated by $\Delta E^{pn}_{I}$. Nevertheless, the reduction of $\alpha
_{c}$
reduces not only $\Delta E^{pn}_{c}$, but also the absolute magnitude of
$E_{c}$ and this
in turn  may jeopardize the fit to the experimental baryon masses. We address
this
problem in the following sections.
\par}
\medskip
{\cwparbegin{\cwfull}{0.00in}{0.00in}{\cwdouble}{0.00in}
{\bf   IV. Instanton-induced mass shifts of baryons}
\par}
{\cwparbegin{\cwfull}{0.00in}{0.00in}{\cwdouble}{0.38in}
\cwindent{0.38in}Unlike for the $p-n$ mass difference, the isospin breaking
does not play a
significant role for the mass shifts (7). We thus take $m_{d}= m_{u}$ so that
$I_{d} = I_{u}$.
The isosymmetric version of the mass shifts due to ${\cal L}^{I}_{1}$ is then
\par}
\medskip
{\cwparbegin{\cwtable}{0.00in}{0.00in}{\cwdouble}{0.38in}
\cwindent{0.38in}\halign{\hfil#\hfil\hskip 3em &\hfil#\hskip 3em \cr
\hskip 2.5cm $\Delta M^{(1)}_{N}= {\widetilde n}\rho ^{-1}4\pi ^{2}{\cal
F}_{u}I_{u}
= \Delta M^{(1)}_{\Delta _{3/2}}$,&(12)\cr
}\par}
\medskip
{\cwparbegin{\cwtable}{0.00in}{0.00in}{\cwsingle}{0.00in}
\halign{\hfil#\hfil\hskip 3em &\hfil#\hskip 3em \cr
\hskip 2.5cm $\Delta M^{(1)}_{\Sigma }= \Delta M^{(1)}_{\Lambda}
={\widetilde n} \rho ^{-1}{4\pi { } ^{2}\over 3}  [2{\cal F}_{u}I_{u}+ {\cal
F}_{s}I_{s}]
= \Delta M^{(1)}_{\Sigma ^{*}_{3/2}}$,&(13)\cr
}\par}
\medskip
\medskip
{\cwparbegin{\cwtable}{0.00in}{0.00in}{\cwsingle}{0.00in}
\halign{\hfil#\hfil\hskip 3em &\hfil#\hskip 3em \cr
\hskip 2.5cm $\Delta M^{(1)}_{\Xi }={\widetilde n}\rho ^{-1}{4\pi { } ^{2}\over
3}
[ {\cal F}_{u} I_{u}+ 2 {\cal F}_{s} I_{s}] = \Delta M^{(1)}_{\Xi
^{*}_{3/2}}$,&(14)\cr
}\par}
\medskip
\medskip
{\cwparbegin{\cwtable}{0.00in}{0.00in}{\cwsingle}{0.00in}
\halign{\hfil#\hfil\hskip 3em &\hfil#\hskip 3em \cr
\hskip 2.5cm \hskip 1.7cm   $  {\widetilde n}\rho^{-1} {4\pi^2}{\cal F}_s  I_s
	     =  \Delta M^{(1)}_{\Omega^-_{3/2}}$. &(15)\cr
}\par}
{\cwparbegin{\cwfull}{0.00in}{0.00in}{\cwdouble}{0.38in}
\cwindent{0.38in}In order to make later some comparisons with the bag model fit
of DeGrand
et al [13], we will quote the numerical results for quark masses they used,
namely $m_{u}= m_{d}= 0,  \,  m_{s}= 280$ MeV:
\par}
\medskip
{\cwparbegin{\cwtable}{0.00in}{0.00in}{\cwdouble}{0.38in}
\cwindent{0.38in}\halign{\hfil#\hfil\hskip 3em &\hfil#\hskip 3em \cr
\hskip 2.5cm $\Delta M^{(1)}_{N}= {\widetilde  n} \cdot  26971$ MeV $=
\Delta M^{(1)}_{\Delta _{3/2}}$, &(16)\cr
}\par}
\medskip
{\cwparbegin{\cwtable}{0.00in}{0.00in}{\cwsingle}{0.00in}
\halign{\hfil#\hfil\hskip 3em &\hfil#\hskip 3em \cr
\hskip 2.5cm $\Delta M^{(1)}_{\Sigma }= \Delta M^{(1)}_{\Lambda}
= {\widetilde n}\cdot  23949$ MeV $= \Delta M^{(1)}_{\Sigma
^{*}_{3/2}}$,&(17)\cr
}\par}
\medskip
\medskip
{\cwparbegin{\cwtable}{0.00in}{0.00in}{\cwsingle}{0.00in}
\halign{\hfil#\hfil\hskip 3em &\hfil#\hskip 3em \cr
\hskip 2.5cm $\Delta M^{(1)}_{\Xi }={\widetilde n} \cdot  20927$ MeV
$= \Delta M^{(1)}_{\Xi ^{*}_{3/2}}$ &(18)\cr
}\par}
\medskip
\medskip
{\cwparbegin{\cwtable}{0.00in}{0.00in}{\cwsingle}{0.00in}
\halign{\hfil#\hfil\hskip 3em &\hfil#\hskip 3em \cr
\hskip 2.5cm \hskip 1.7cm $ {\widetilde n} \cdot 17905$ MeV
= $\Delta M^{(1)}_{\Omega^-_{3/2}}$. &(19)\cr
}\par}
{\cwparbegin{\cwfull}{0.00in}{0.00in}{\cwdouble}{0.00in}
Results for $m_{u}= m_{d}= 8$ MeV and the more standard
strange mass $m_{s}= 200$ MeV
differ from $(16)-(19)$ only slightly, by 2\% to 5\%.
\par}
{\cwparbegin{\cwfull}{0.00in}{0.00in}{\cwdouble}{0.38in}
\cwindent{0.38in}The mass shifts due to the two-body term are given
basically by a simple integral over the square of the sum of
squared Bessel functions only for the
(isosymmetric) nucleons, with $m_u =  m_d$.
For the $\Sigma$,  $\Xi$ and $\Lambda$, the
corresponding expressions are more involved  because the significantly
different  mass  of  the strange quark complicates them slightly:
$$ \Delta M_N^{(2)} = - 9 {\cal K}_{ud} \left( {{\cal  N}^2\over 8\pi
}\right)^2
\int_{V_{bag}}  ( A_+^2  j_0^2  +  A_-^2  j_1^2 )^2  \,  d^3r  ,   \qquad
\qquad  (20)  $$
\medskip
$$  \Delta M_\Sigma^{(2)}  =  \Delta M_\Xi^{(2)}  =
 -  {\cal K}_{us} {{\cal N}^2 {\cal N}_s^2  \over  (8\pi)^2 }  \int_{V_{bag}}
\lbrace  9 (A_+^2  j_0^2  +  A_-^2  j_1^2 )
 (S_+^2 {\widetilde j}_0^2  +  S_-^2 {\widetilde j}_1^2)  \qquad $$
$$\hskip 1cm \qquad\qquad \qquad  - 10 (A_+  j_0  S_-  {\widetilde j}_1
-  A_-  j_1  S_+ {\widetilde j}_0)^2
\rbrace  \,   d^3r ,  \qquad \qquad  \hskip 1cm  (21)   $$
\medskip
$$  \Delta  M_\Lambda^{(2)}  =
		     - 6  {\cal K}_{ud} \left( { {\cal  N}^2\over 8\pi }\right)^2
		     \int_{V_{bag}}  ( A_+^2  j_0^2  +  A_-^2  j_1^2 )^2  \,  d^3r
  -  3  {\cal K}_{us} {{\cal N}^2 {\cal N}_s^2  \over  (8\pi)^2 }      \qquad
$$
$$\qquad
\hskip  1cm  \times \int_{V_{bag}}
 \left[ ( A_+  j_0 S_+ {\widetilde j}_0  + A_-  j_1 S_ -{\widetilde j}_1 )^2
  - (  A_-  j_1 S_+ {\widetilde j}_0  -  A_+ j_0  S_-  {\widetilde j}_1 )^2
\right]
		                                                         \,   d^3r  . \qquad
(22)   $$
$j_i$ and  $ {\widetilde  j}_i$  ($i=0,1$) are  the  $i$-th spherical
Bessel functions of the arguments  $x_u  r/R_{bag}$ and  $x_s r/R_{bag}$,
respectively:
$$ j_i \equiv  j_i(x_u  r/R), \qquad  {\widetilde j}_i  \equiv  j_i (x_s r/R)
,\qquad  (i=0,1)  \qquad (23)   $$
where  $x_f$  is the bag eigenvalue  of  the lowest  mode  of the quark of  the
mass
$m_f$  ($f=u,d,s$).  The energy  of the quark of the flavor $f$ is
correspondingly
$$ \omega_f  =  \sqrt{ x_f^2 /R_{bag}^2 +  m_f ^2 }.    $$
We  have used  the  following  abbreviations:
$$  A_{\pm}^2 \equiv  {\omega_u  \pm  m_u  \over  \omega_u},   \qquad\quad
      S_{\pm}^2  \equiv  {\omega_s  \pm   m_s  \over  \omega_s},   \qquad \quad
 (24) $$
$$  {\cal K}_{f_1 f_2} =  n  \left({4\pi^2 \over  3}  \rho^3 \right)^2
	       {\cal F}_{f1} {\cal F}_{f_2},\qquad     f = u,d,s.    \qquad  (25)  $$
Now we are in the isosymmetric limit,  so that $x_u  =  x_d$,
$\omega_u  =  \omega_d$  and   ${\cal F}_u =  {\cal F}_d$.
${\cal N}$  is  the  usual normalization for $u$ and $d$, and
${\cal N}_s$ for $s$ quark wavefunctions.

Here we quote the numerical values for $m_{u}= m_{d}= 0, m_{s}= 280$ MeV:
\par}
\medskip
{\cwparbegin{\cwtable}{0.00in}{0.00in}{\cwdouble}{0.00in}
\halign{\hfil#\hfil\hskip 3em &\hfil#\hskip 3em \cr
\hskip 3.2cm $\Delta M^{(2)}_{N}= - {\widetilde n} \cdot  12468 $ MeV&(26)\cr
}\par}
\medskip
{\cwparbegin{\cwtable}{0.00in}{0.00in}{\cwsingle}{0.00in}
\halign{\hfil#\hfil\hskip 3em &\hfil#\hskip 3em \cr
\hskip 3.2cm $\Delta M^{(2)}_{\Sigma }= \Delta M^{(2)}_{\Xi }=
- {\widetilde n} \cdot $ 6341 MeV&(27)\cr
}\par}
\medskip
\medskip
{\cwparbegin{\cwtable}{0.00in}{0.00in}{\cwsingle}{0.00in}
\halign{\hfil#\hfil\hskip 3em &\hfil#\hskip 3em \cr
\hskip 3.2cm $\Delta M^{(2)}_{\Lambda}= - {\widetilde n}\cdot 10413 $
MeV.&(28)\cr
}\par}
\medskip
{\cwparbegin{\cwfull}{0.00in}{0.00in}{\cwdouble}{0.00in}
(For moderately different masses, e.g. $m_{u}= m_{d}= 8$ MeV and $m_{s}= 200$
MeV, results
again differ only slightly).
\par}
{\cwparbegin{\cwfull}{0.00in}{0.00in}{\cwdouble}{0.38in}
 \cwindent{0.38in}For the baryons from the decuplet, the ${\cal L}^{I}_{2}$
 contribution vanishes, $\Delta M^{(2)}_{decuplet}= 0$. The total instanton
contribution $E^{B}_{I}\equiv  \Delta M^{(1)}_{B}+ \Delta M^{(2)}_{B}$
is thus (for $m_{u}= m_{d}=0,  m_{s}= 280$ MeV and $R = 5$ GeV$^{-1})$:
\par}
\medskip
{\cwparbegin{\cwtable}{0.00in}{0.00in}{\cwdouble}{0.38in}
\cwindent{0.38in}\halign{\hfil#\hfil\hskip 3em &\hfil#\hskip 3em \cr
\hskip 3.2cm $E^{N}_{I}= {\widetilde n}\cdot  14503 $  MeV,&(29a)\cr
}\par}
\medskip
{\cwparbegin{\cwtable}{0.00in}{0.00in}{\cwsingle}{0.00in}
\halign{\hfil#\hfil\hskip 3em &\hfil#\hskip 3em \cr
\hskip 3.2cm $E^{\Sigma }_{I}= {\widetilde n}\cdot 17608 $ MeV,&(29b)\cr
}\par}
\medskip
{\cwparbegin{\cwtable}{0.00in}{0.00in}{\cwsingle}{0.00in}
\halign{\hfil#\hfil\hskip 3em &\hfil#\hskip 3em \cr
\hskip 3.2cm $E^{\Xi }_{I}= {\widetilde n} \cdot 14631 $  MeV,&(29c)\cr
}\par}
\medskip
{\cwparbegin{\cwtable}{0.00in}{0.00in}{\cwsingle}{0.00in}
\halign{\hfil#\hfil\hskip 3em &\hfil#\hskip 3em \cr
\hskip 3.2cm $E^{\Delta _{3/2}}_{I}= {\widetilde n} \cdot 26971 $
MeV,&(29d)\cr
}\par}
\medskip
{\cwparbegin{\cwtable}{0.00in}{0.00in}{\cwsingle}{0.00in}
\halign{\hfil#\hfil\hskip 3em &\hfil#\hskip 3em \cr
\hskip 3.2cm $E^{\Sigma ^{*}_{3/2}}_{I}= {\widetilde n}\cdot 23949 $
MeV,&(29e)\cr
}\par}
\medskip
{\cwparbegin{\cwtable}{0.00in}{0.00in}{\cwsingle}{0.00in}
\halign{\hfil#\hfil\hskip 3em &\hfil#\hskip 3em \cr
\hskip 3.2cm $E^{\Xi ^{*}_{3/2}}_{I}= {\widetilde n}\cdot 20927 $ MeV,&(29f)\cr
}\par}
\medskip
{\cwparbegin{\cwtable}{0.00in}{0.00in}{\cwsingle}{0.00in}
\halign{\hfil#\hfil\hskip 3em &\hfil#\hskip 3em \cr
\hskip 3.2cm $E^{\Omega ^{-}_{3/2}}_{I}= {\widetilde n}\cdot 17905 $
MeV.&(29g)\cr
}\par}
\medskip
{\cwparbegin{\cwfull}{0.00in}{0.00in}{\cwdouble}{0.00in}
($\Lambda$ is left out here, since  we would need also
the three-body piece to form $E^{\Lambda}_{I}$, the
total instanton contribution  for  this  baryon.  The sum of one- and two-body
instanton contributions  for $\Lambda$ is
$\Delta M^{(1)}_{\Lambda}+ \Delta M^{(2)}_{\Lambda}={\widetilde n} \cdot 13536
$  MeV.)
\par}
\medskip
\noindent {\bf   V.  Estimating the instanton density} ${\widetilde n}$
{\bf  inside the bag}
{\cwparbegin{\cwfull}{0.00in}{0.00in}{\cwdouble}{0.38in}
\cwindent{0.38in}
\medskip

Determining the value of
${\widetilde n}$  consistent with the MIT bag interior also means
exploring the possibility of reducing the value of $\alpha _{c}$,
which would improve the
consistency of the perturbative approach inside the bag. This is how we can
accommodate $\Delta E^{pn}_{I}$, i.e. our result for instanton contribution to
the proton-neutron mass difference. The decrease of $\alpha _{c}$
would reduce $\Delta E^{pn}_{c}$ and this reduction
would be compensated by $\Delta E^{pn}_{I}$. However,
the reduction of $\alpha _{c}$  decreases not only
$\Delta E^{pn}_{I}$, but also the absolute magnitude of the
chromomagnetic energy $E_{c}$. Since $E_{c}$
is negative for nucleons and other octet baryons, it will not be possible to
compensate by the total instanton contribution $E_{I}$ the decrease in the
absolute
value of  $E_{c}$ resulting from the decrease of $\alpha _{c}$. This is because
$E_{I}$  is {\bf positive}
since the one-body contribution exceeds the two-body one. The nucleon-delta
splitting would also be spoiled for the same reasons.
\par}
{\cwparbegin{\cwfull}{0.00in}{0.00in}{\cwdouble}{0.38in}
\cwindent{0.38in}Still, the decrease in $\alpha _{c}$, along the good fit to
the hadron masses, may be
achieved anyway if, along the inclusion of $E_{I}$, other contributions to the
bag
energy are also changed. (For example the zero-point energy $-Z_{0}/R$ is
negative
and the increase in the parameter $Z_{0}$ may compensate for the
decrease of $\alpha _{c}$ and
the inclusion of $E_{I}$. If we start from, e.g., the bag model fit of De Grand
et
al.  [13], where the bag model parameters were chosen so that they reproduce
the
experimental nucleon mass $M_{N}$, then we demand that after the inclusion of
the
instanton contribution $E_{I}$  (that is, allowing  ${\widetilde n}$ inside the
bag
to deviate from   $ \widetilde n = 0$ ), the bag
model parameters change so that the nucleon mass remains at the empirical
value:
\par}

$$\delta M_{N} \!  = \! {\delta M_{N}\over \delta R_{N}} \delta R_{N} \! +
\! {\delta M_{N}\over \delta \alpha_{c}} \delta \alpha_{c} \! +
\! {\delta M_{N}\over \delta Z_{0}} \delta Z_{0}\! +
\! {\delta M_{N}\over \delta B} \delta B
\! + \! {\delta M_{N}\over \delta {\widetilde n}} \delta {\widetilde n} \! = \!
0 . \qquad  (30) $$

\medskip
{\cwparbegin{\cwfull}{0.00in}{0.00in}{\cwdouble}{0.38in}
\cwindent{0.38in}
\noindent (We do not consider the possibility of varying the quark masses
because
we have to adopt the values used by Ref. [13] if we want to use the results of
Refs.
[13]  and [17], and to compare our results with theirs. Bickerstaff and Thomas
[17] themselves studied the effects of small and different $u$ and $d$ masses
on top
of the isosymmetric fit of De Grand et al.  [13].)
\par}
{\cwparbegin{\cwfull}{0.00in}{0.00in}{\cwdouble}{0.38in}
\cwindent{0.38in}
Fortunately, it is not necessary to perform a full refitting of the bag-
model parameters varied in Eq. (30) as it is a good approximation to "freeze"
the bag radii because of the pressure-balance condition
\par}
$${dM_{N}\over dR_{N}} = 0.   \qquad\qquad\qquad\qquad  (31) $$
{\cwparbegin{\cwfull}{0.00in}{0.00in}{\cwdouble}{0.00in}
Namely, we have seen above that since $p-n$ mass difference is small,
so that there is room only for relatively small instanton effects. More
precisely, even if almost the whole of $\Delta E^{pn}_{c}$ is supplanted
by $\Delta E^{pn}_{I}$, the maximal $\widetilde n$
that can be accommodated by $\Delta E_{pn}$ would be
$\widetilde n  \approx  0.3\cdot 10^{-2}$, so that $E^{N}_{I}$cannot be
excessively large. Also, we commented above that the conflicting behaviours of
$\Delta E^{pn}_{I}, E_{I}$ and $E_{c}$ with $\widetilde n$ and $\alpha _{c}$
cannot allow large instanton effects. Therefore,
the fit of Ref.  [13] cannot be altered very much,
so that the pressure-balance condition  (31)
still holds in a good approximation after the inclusion of $E_{I}$. Thus, the
variation of the radius $R_{N}$ cannot contribute to the energy-variation
equation
(30) comparably to the variations of other parameters, i.e., we can neglect the
term
\par}
$${\delta M\over \delta R} \delta  R. $$
\medskip
{\cwparbegin{\cwfull}{0.00in}{0.00in}{\cwdouble}{0.00in}
Only very large variations of $R$ would change the energy balance
substantially. (Indeed, a change of $R$ by as much as 10\% changes
the mass only by order of 1\%.)
This is a significant simplification of Eq. $(30)$ since the remaining
parameters
$(Z_{0}, B, \alpha_{c}, n)$ enter linearly.
\par}
{\cwparbegin{\cwfull}{0.00in}{0.00in}{\cwdouble}{0.38in}
\cwindent{0.38in}We therefore consider only the change
of $Z^{old}_{0}, \alpha ^{old}_{c}$ and $B_{old}$
(where "$old$" indicates the values of Ref. [13], {\it  i.e.} 1.84, 0.55 and
(0.145 GeV)$^{4}$, respectively) to new values $Z_{0}, \alpha_{c}$ and
$B$ when the instanton induced interaction is turned on,
i.e.  $\widetilde n$ allowed to deviate from zero. Eq. (30)
then leads (in the case of nucleon $N$) to
\par}

$$M^{bag}_{N} - E^{N}_{Q} = {Z_{0}\over Z^{old}_{0}} E^{N}_{0} +
{\alpha_{c}\over \alpha^{old}_{c}} E^{N}_{M}+ {B\over B_{old}} E_{V}+E_{I} ,
\qquad\qquad (32)$$
{\cwparbegin{\cwfull}{0.00in}{0.00in}{\cwdouble}{0.00in}
where $E^{N}_{Q}, E^{N}_{M}, E^{N}_{V}$and $E^{N}_{0}$ are,
respectively, the kinetic, chromomagnetic, volume,
and zero-point energies for the nucleon ($N$) as given in
Table III of De Grand et al.  [13]. Ref.  [13] also fits the
mass of the $\Omega^{-}_{3/2}$  bag to the experimental value,
just as for the nucleon and delta bags.
Since we are interested in the interplay of instanton effects
and one-gluon exchange, we are especially concerned with not
spoiling nucleon-delta splitting, which is usually attributed
to one-gluon exchange. We thus demand that equations
analogous to the equation (30) for the nucleon, or,
equivalently, (32), also hold for
$\Delta_{3/2}$ and $\Omega ^{-}_{3/2}$. For the fourth equation
necessary for determining the four unknowns (the new $Z_{0}$,  $\alpha_{c}$
and $B$, but also $\widetilde n$  sitting in $E_{I}$ and $\Delta E^{np}_{I})$
we have chosen the equation for $p-n$ mass difference in the
presence of instanton-induced interaction ${\cal L}_{I}$:
\par}
$$\Delta E_{pn} = \Delta E^{pn}_{K} + \Delta E^{pn}_{EM} +
{\alpha_{c}\over \alpha^{old}_{c}} \Delta E_{c} + \Delta E^{pn}_{I} .
\qquad\qquad  (33)$$
{\cwparbegin{\cwfull}{0.00in}{0.00in}{\cwdouble}{0.00in}
Why this choice? As solving of the equation (33) for
the instanton density $\widetilde n$ shows,
(33) ensures the positivity of
$\widetilde n$ if the expected decrease of $\alpha_c$
indeed occurs. (And it does, as we will see below.) In addition
to the discussion in Section III, which indicates that
$p-n$ mass difference might hold an important message
for instantons in the M.I.T. bag, the issue of positivity
of the instanton density
(proportional to the number of instantons
{\it plus} the number of anti-instantons)
was the other reason for
choosing (33) for the fourth determining equation. Namely, we
could have of course also taken an equation analogous to the
aforementioned  equations for $N$, $\Delta$ or $\Omega$,
but for a baryon other than these already used. In that
case, however, there is nothing to act automatically against
the solutions with negative instanton density, and this
physical requirement would have to be imposed as an additional
constraint. (Indeed, if we use $\Xi$ or $\Sigma$, it turns
out that, although the results for $\alpha_c$
and the {\it absolute} magnitude of
instanton effects are qualitatively similar
as when $p-n$ mass difference is used,
the instanton density comes out positive
in the case of the $\Xi$-equation, but negative
in the case of the fourth equation stemming from
$\Sigma$.)
\par}
{\cwparbegin{\cwfull}{0.00in}{0.00in}{\cwdouble}{0.00in}
In (33), we have used the values of  Ref. [17] for $\Delta E_{K},
\Delta E_{EM}, \Delta E_{c}$ and $\Delta E_{pn}$.
For the quark masses used by
Ref.  [17] ($m_{u} = 7.86$ MeV, $m_{d}= 12.14$ MeV),
\par}
$$\Delta E^{pn}_{I}=  {\widetilde n} \cdot 83.14
\,\,  {\rm MeV}.    \qquad\qquad\qquad  (34) $$
\medskip
{\cwparbegin{\cwfull}{0.00in}{0.00in}{\cwdouble}{0.00in}
We make the same approximation as that used
by Bickerstaff and Thomas [17], i.e. we use the energy
differences $\Delta E^{pn}_{X}$  $(X = K, EM, c, I)$
computed for the small and different quark
masses $m_{u}, m_{d}$ on top of the isosymmetric
fit to hadron masses done in Ref. [13]
with $m_{u}= m_{d}= 0,  m_{s}= 280$ MeV.
\par}
{\cwparbegin{\cwfull}{0.00in}{0.00in}{\cwdouble}{0.38in}
\cwindent{0.38in}  Solving  the system of four linear equations
consisting of  (33), (32)
and analogies of (32)  for $\Omega^-_{3/2} $ and for $\Delta_{3/2} $,
yields $\alpha _{c}= 0.52, Z_{0}= 2.11, B = (0.151$ GeV$)^{4}$ and
$ {\widetilde n} = 0.205\cdot 10^{-3}$. This  ${\widetilde n}$
corresponds to $n = (71.79$ MeV$)^{4}$ for
$\rho  = {1/ 600\hbox{ MeV}}$ and is even more depleted
with respect to the $n$ in the nonperturbative QCD vacuum than we
expected. (E.g., it is roughly ${1/30}$ of Shuryak's estimate  [12]
for nonperturbative vacuum.)
\par}
{\cwparbegin{\cwfull}{0.00in}{0.00in}{\cwdouble}{0.38in}
\cwindent{0.38in} For $m_{u}= m_{d}= 0$ and $m_{s}= 280$ MeV,
i.e. the values used in Table III of Ref. [13], the instanton
contribution to the energies for the octet baryons are
\par}
{\cwparbegin{\cwfull}{1.23in}{0.00in}{\cwdouble}{0.00in}
$E^{N}_{I}= 3.0$ MeV,
\par}
{\cwparbegin{\cwfull}{1.23in}{0.00in}{\cwdouble}{0.00in}
$E^{\Sigma }_{I}= 3.6$ MeV,
\par}
{\cwparbegin{\cwfull}{1.23in}{0.00in}{\cwdouble}{0.00in}
$E^{\Xi }_{I}= 3.0$ MeV,
\par}
\medskip
{\cwparbegin{\cwfull}{0.00in}{0.00in}{\cwdouble}{0.00in}
(as for $\Lambda $,  $\Delta M^{(1)}_{\Lambda }+ \Delta M^{(2)}_{\Lambda }=
2.8$ MeV),
 and for the decuplet baryons,
\par}
\medskip
{\cwparbegin{\cwfull}{1.23in}{0.00in}{\cwdouble}{0.00in}
$E^{\Delta_{3/2} }_{I}= 5.5$ MeV,
\par}
{\cwparbegin{\cwfull}{1.23in}{0.00in}{\cwdouble}{0.00in}
$E^{\Sigma ^{*}_{3/2}}_{I}= 4.9$ MeV,
\par}
{\cwparbegin{\cwfull}{1.23in}{0.00in}{\cwdouble}{0.00in}
$E^{\Xi ^{*}_{3/2}}_{I}= 4.3$ MeV,
\par}
{\cwparbegin{\cwfull}{1.23in}{0.00in}{\cwdouble}{0.00in}
$E^{\Omega ^{-}_{3/2}}_{I}= 3.7$ MeV.
\par}
{\cwparbegin{\cwfull}{0.00in}{0.00in}{\cwdouble}{0.00in}
The results for $m_{u}= m_{d}= 8$ MeV and $m_{s}= 200$ MeV are very similar.
\par}
{\cwparbegin{\cwfull}{0.00in}{0.00in}{\cwdouble}{0.38in}
\cwindent{0.38in}With these instanton-induced additions, and with the
changed $E_{0}, E_{V}$ and $E_{M}$ contributions (while $E_{Q}$ and
$E_{E}$ stay as they are in Table III of Ref. [13] ), the bag masses
are for the octet
\par}
{\cwparbegin{\cwfull}{1.23in}{0.00in}{\cwdouble}{0.00in}
$M^{N}_{bag}= M^{N}_{\exp}= 938$ MeV,
\par}
{\cwparbegin{\cwfull}{1.23in}{0.00in}{\cwdouble}{0.00in}
$M^{\Sigma }_{bag}= 1142$ MeV,
\par}
{\cwparbegin{\cwfull}{1.23in}{0.00in}{\cwdouble}{0.00in}
$M^{\Xi }_{bag}= 1285$ MeV,
\par}
\medskip
{\cwparbegin{\cwfull}{0.00in}{0.00in}{\cwdouble}{0.00in}
(for $\Lambda $, modulo ${\cal L}^{I}_{3}$ contribution,
$M^{\Lambda }_{bag}= 1103$ MeV), and for the decuplet:
\par}
\medskip
{\cwparbegin{\cwfull}{1.23in}{0.00in}{\cwdouble}{0.00in}
$M^{\Delta_{3/2} }_{bag}= M^{\Delta_{3/2} }_{\exp}= 1233$ MeV,
\par}
{\cwparbegin{\cwfull}{1.23in}{0.00in}{\cwdouble}{0.00in}
$M^{\Sigma ^*_{3/2}}_{bag}= 1385$ MeV,
\par}
{\cwparbegin{\cwfull}{1.23in}{0.00in}{\cwdouble}{0.00in}
$M^{\Xi ^{*}_{3/2}}_{bag}= 1529$ MeV,
\par}
{\cwparbegin{\cwfull}{1.23in}{0.00in}{\cwdouble}{0.00in}
$M^{\Omega^-_{3/2} }_{bag}= M^{\Omega ^-_{3/2}}_{\exp}= 1672$ MeV.
\par}
\medskip
{\cwparbegin{\cwfull}{0.00in}{0.00in}{\cwdouble}{0.00in}
The fit to the baryon masses therefore remains good.
\par}
{\cwparbegin{\cwfull}{0.00in}{0.00in}{\cwdouble}{0.38in}
\cwindent{0.38in}

\medskip

{\noindent \bf   VI.   Conclusion}

We have found that at least in the simplest, MIT  bag model, the
instanton-induced baryon mass shifts are not similar  to the one-gluon
exchange effects (unlike the situation  found in the nonrelativistic
constituent
quark model [1]). The instanton-induced mass shifts in the MIT bag model are
small, of the order of a few MeV,  just as the reduction of the strong coupling
constant is just about 6\%. That is, the change in $\alpha _{c}$ is in the
desired
direction, but it is quantitatively so marginal that we cannot claim that the
inclusion of the instanton effects has successfully supplanted  or supplemented
the one-gluon exchange and has brought about the desired improvement in
consistency of the perturbative description of the bag interior. It has turned
out, however, that the "perturbative" bag interior cannot support
sufficiently high instanton density for ensuring an important role
of instantons in the MIT bag model. The value of $\widetilde n$
we have estimated as appropriate for the MIT bag interior is one to
two orders of magnitude lower than $\widetilde n$ estimated e.g.
by Nowak et al. [6]  or Shuryak [12], but for the true,
nonperturbative QCD vacuum. Such instanton densities would
induce gigantic mass shifts of hundreds of MeV.
At a conceptual level, this actually shows us another reason why the MIT bag
model cannot tolerate high instanton densities. If $\widetilde n$ is too high,
the "dressing"
of quarks would also be large, and a too large quark self-mass would lead to
{\it de
facto} constituent, nonrelativistic quarks, while the MIT bag is a relativistic
model. Therefore, instanton densities above a certain value would lead to a
contradiction with the relativistic way the model was formulated. The
relativistic nature of the MIT bag model, which requires the inclusion of the
one-body of the instanton-induced interaction, is the reason why including of
instantons gives different results than in the nonrelativistic constituent
model [1] .
\par}
{\cwparbegin{\cwfull}{0.00in}{0.00in}{\cwdouble}{0.38in}
\cwindent{0.38in}Finally, let us remark that there are physical situations
where even the
weak instanton interaction, which produces unimportant effects on MIT bag model
spectroscopy, can have interesting effects. Production of strangeness in
nucleons
and some of its consequences is one such example. The work on these issues is
in
progress.
\par}
\medskip
\vskip 1cm
{\cwparbegin{\cwfull}{0.00in}{0.00in}{\cwdouble}{0.00in}
{\bf Acknowledgement:} I thank  Dr. I. Zahed  for  pointing
out this problem to me, and both to him and Dr. M. Nowak for
illuminating  discussions  during  my  visits to  Nuclear Theory Group  at
Physics Dept. of  Stony Brook University,  and  especially to NSF which
made  these visits  and this work possible  by supporting  it  through  the
contract JF 899 - 31. The partial support of the EC contract
CI1*-CT91-0893(HSMU) is also acknowledged.
\par}
\medskip
\newpage
{\cwparbegin{\cwfull}{0.00in}{0.00in}{\cwdouble}{0.00in}
{\bf References:}
\par}
{\cwparbegin{\cwfull}{0.31in}{0.00in}{\cwdouble}{-0.31in}
\cwoutdent{0.31in}{~1.} E. Shuryak and Rosner, Phys. Lett. {\bf  218B}  (1989)
72.
\par}
{\cwparbegin{\cwfull}{0.31in}{0.00in}{\cwdouble}{-0.31in}
\cwoutdent{0.31in}{~2.} A. Chodos,  R.L. Jaffe, C.B.  Thorn
and V. Weiskopf, Phys. Rev. {\bf D9 }  (1974)  3471 ;  A. Chodos, R. L. Jaffe,
K. Johnson and C. B. Thorn,  Phys. Rev.  {\bf D10}  (1974)  2599.
\par}
{\cwparbegin{\cwfull}{0.31in}{0.00in}{\cwdouble}{-0.31in}
\cwoutdent{0.31in}{~3.} A. Chodos and C.B. Thorn, Phys. Rev. {\bf D12} (1975)
2733.
\par}
{\cwparbegin{\cwfull}{0.31in}{0.00in}{\cwdouble}{-0.31in}
\cwoutdent{0.31in}{~4.} A  very good review is G.E. Brown and M. Rho,
Comments on Nucl. and Part. Phys. {\bf 18} (1988) 1.
\par}
{\cwparbegin{\cwfull}{0.31in}{0.00in}{\cwdouble}{-0.31in}
\cwoutdent{0.31in}{~5.} J. Greensite, Nucl. Phys. {\bf B249} (1985) 263;
Yu. A.  Simonov, Yad. Fiz. {\bf 50}  (1989) 500.
\par}
{\cwparbegin{\cwfull}{0.31in}{0.00in}{\cwdouble}{-0.31in}
\cwoutdent{0.31in}{~6.} M.A. Nowak, J.J.M. Verbaarschot and
I. Zahed, Nucl. Phys. {\bf B324} (1989) 1.
\par}
{\cwparbegin{\cwfull}{0.31in}{0.00in}{\cwdouble}{-0.31in}
\cwoutdent{0.31in}{~7.} M.A. Shifman, A.I. Vainshtein and
V.I. Zakharov, Nucl. Phys. {\bf B163} (1980) 46.
\par}
{\cwparbegin{\cwfull}{0.31in}{0.00in}{\cwdouble}{-0.31in}
\cwoutdent{0.31in}{~8.} M.A. Nowak, Acta Physica Polonica {\bf B22} (1991) 697.
\par}
{\cwparbegin{\cwfull}{0.31in}{0.00in}{\cwdouble}{-0.31in}
\cwoutdent{0.31in}{~9.} N.I. Kochelev, Yad. Fiz. {\bf 41} (1985) 456.
\par}
{\cwparbegin{\cwfull}{0.31in}{0.00in}{\cwdouble}{-0.31in}
\cwoutdent{0.31in}{10.} N.I. Kochelev, Yad. Fiz. {\bf 39} (1984) 462.
\par}
{\cwparbegin{\cwfull}{0.31in}{0.00in}{\cwdouble}{-0.31in}
\cwoutdent{0.31in}{11.} C. Callan, R. Dashen and D. Gross,
Phys. Lett. ${\bf 7}{\bf 8}{\bf B} (1978) 307$;
R.L. Jaffe, Lectures at Erice Summer School 1979;
G.E. Brown and M. Rho, Phys. Lett. ${\bf 8}{\bf 2}{\bf B}
(1979) 177$ and Ref. [4] above;
A.W. Thomas, Adv. Nucl. Phys. ${\bf 1}{\bf 3} (1983) 1$.
\par}
{\cwparbegin{\cwfull}{0.31in}{0.00in}{\cwdouble}{-0.31in}
\cwoutdent{0.31in}{12.} E.V. Shuryak, Nucl. Phys. ${\bf B}{\bf 2}{\bf 0}{\bf 3}
(1982) 93$.
\par}
{\cwparbegin{\cwfull}{0.31in}{0.00in}{\cwdouble}{-0.31in}
\cwoutdent{0.31in}{13.} T. DeGrand, R.L. Jaffe, K. Johnson
and J. Kiskis, Phys. Rev. ${\bf D}{\bf 1}{\bf 2} (1975)
2060$.
\par}
{\cwparbegin{\cwfull}{0.31in}{0.00in}{\cwdouble}{-0.31in}
\cwoutdent{0.31in}{14.} D.I. Dyakonov and V.Yu. Petrov,
Nucl. Phys. ${\bf B}{\bf 2}{\bf 4}{\bf 5} (1984) 256$, and Nucl. Phys.
${\bf B}{\bf 2}{\bf 7}{\bf 2} (1986) 457$;
E.V. Shuryak, Nucl. Phys. ${\bf B}{\bf 3}{\bf 0}{\bf 2} (1988) 599$.
\par}
{\cwparbegin{\cwfull}{0.31in}{0.00in}{\cwdouble}{-0.31in}
\cwoutdent{0.31in}{15.} A. Chodos and C. B. Thorn, Nucl. Phys. {\bf  B104}
(1989)  72.
\par}
{\cwparbegin{\cwfull}{0.31in}{0.00in}{\cwdouble}{-0.31in}
\cwoutdent{0.31in}{16.} N.G. Deshpande, D.A. Dicus, K. Johnson
and V.L. Teplitz, Phys. Rev. {\bf D15} (1977) 1885.
\par}
{\cwparbegin{\cwfull}{0.31in}{0.00in}{\cwdouble}{-0.31in}
\cwoutdent{0.31in}{17.} R.P. Bickerstaff and A.W. Thomas, Phys. Rev. {\bf D 25}
(1982) 1869.
\par}
{\cwparbegin{\cwfull}{0.31in}{0.00in}{\cwdouble}{-0.31in}
\cwoutdent{0.31in}{18.} G.A. Miller, B.M.K. Nefkens and I. \v Slaus,
Phys. Rep. {\bf 194}, no. 1. \& 2. (1990).
\par}

\end{document}